\documentclass[12pt]{article}
\usepackage{epsfig}
\usepackage{supertabular}

\begin{document}

\title{Broadcasting but not receiving: density dependence considerations for SETI signals}
\author{Reginald D. Smith \\ Bouchet-Franklin Institute \\ P.O. Box 10051 \\ Rochester,
NY 14610, USA \\ \texttt{rsmith@bouchet-franklin.org}}

\maketitle

\begin{abstract}
This paper develops a detailed quantitative model which uses the
Drake equation and an assumption of an average maximum radio
broadcasting distance by an communicative civilization. Using this
basis, it estimates the minimum civilization density for contact
between two civilizations to be probable in a given volume of space
under certain conditions, the amount of time it would take for a
first contact, and the question of whether reciprocal contact is
possible.
\end{abstract}

\section{Introduction}

The question of the existence of extraterrestrial life has long been
one of the most important questions facing mankind. It has inspired
countless books, movies, subcultures, cults, and scientific (or
pseudoscientific) investigations to probe its validity. The most
well-funded and scientifically credible of these efforts is the
ongoing Search for Extraterrestrial Intelligence (SETI).
This program involves the use of both radio and optical
telescopes to search the cosmos for anomalous signals that could
herald the first confirmed contact with another intelligence.

Whether this will happen in our lifetimes, if ever, will continue to
be an uncertain question. How do we know anyone is out there? The
famous question posed by Enrico Fermi \cite{fermi}, otherwise known
as the Fermi Paradox, still begs for an answer. Given our current
technology and reach, the highest level of certainty can only be
obtained by educated conjecture.  Frank Drake, in a lecture at the
first SETI symposium in 1961, came up with the well-known Drake
Equation \cite{drake}. This equation was not really meant to
be an exact mathematical certainty but rather a starting point on an
agenda on the topic how to make an educated guess about the current
number of intelligent, radio-wave communicating species in the
galaxy. The standard Drake Equation is expressed as

\begin{equation}
\label{drake1} N = R^{*}*f_s*L
\end{equation}

where $ R^{*}$ is the average production rate for stars `suitable'
for planets and eventually intelligent life, $f_s$ is the
probability of the emergence of an intelligent and communicating
civilization around one such star, and $L$ is the average lifetime
of such a communicating civilization (hereafter known as CC). In
many representations of the Drake equation, $f_s$ is expanded to
several terms to represent the various probabilities in the
emergence of intelligent, communicating life so for example, in
Shklovskii and Sagan \cite{sagan}, $f_s$ is expressed as

\begin{equation}
\label{drake2} f_s = f_p*n_e*f_L*f_i*f_c
\end{equation}

where $f_p$ is the fraction of suitable stars with planets, $n_e$ is
the average number of habitable (usually assumed to be
Earth-like) planets around each star, $f_L$ is the probability of
life developing on such a planet, $f_i$ is the probability of
intelligent life, and $f_c$ is the probability of intelligent life
developing a technological CC. It should also be noted that the
Drake equation is in fact, an example of a form of the common
equation Little's Law, $N = rt$ where $N$ is the average number of
items in a system, $r$ is the average arrival rate, and $t$ is the
average time in the system. For Drake's equation $r = R^{*}*f_s$ and
$t = L$. The Drake equation, which assumes fixed values for its
parameters, has been adjusted in several works \cite{drakeadvanced,
drakeadvancedb} to take into account possible statistical
distributions for its parameters to allow greater flexibility.

The Drake equation is well-known, but not without its valid
criticisms. In its most common and basic form, it assumes
isotropic conditions across space for stellar formation and
habitable planets where inhomogeneity is the rule and not the
exception. The only quantity which we can currently measure with any
certainty is the star formation rate, $R^*$. Finally, widely varying
and over optimistic or pessimistic estimates of the key parameters
can lead to wildly high or low probabilities for life in our
galactic neighborhood \cite{saganrebut,saganrebutb}.

However, the Drake equation assumes that if intelligent and
technologically advanced life does coexist in our galactic
`neighborhood' we should look and expect to find its tell-tale
signature in the galactic radio noise. However, how should
`neighborhood' be defined? In addition, even if CCs coexist, what if
their signals are faint due to power constraints or distance to the
point  that their neighbors cannot properly distinguish them from
background noise? In fact, given a universe following Drake's Law
for the emergence of intelligent life, how likely is it for a CC
given constraints of survival time and distance where the signals
can reasonably be received, detects a signal from another CC?

The purpose of this paper is to pose basic questions that should,
given appropriate limits, estimate how likely contact is for any
given CC given its mean lifetime $L$ and the mean maximum distance
$D$ a CC's signals can clearly be received.  A common assumption is
if Drake's Equation, $N$, is greater than one, or perhaps in the
range of millions for our galaxy \cite{sagan}, then we should expect
to eventually receive signals from another CC (or vice versa).
Assuming that signals can be detected irrespective of their distance
from their origin, this is a reasonable estimate. However, what if
there is a reasonable horizon for the detection of a signal from
another CC?

\section{A signal through space: volume*times}

The basis of this analysis will rely on the assumption of a CC
around a star which broadcasts for its entire lifetime, $L$. After
time $L$, the CC goes extinct but its signals carry on throughout
space until a distance, $D$ and a time $D/c$, where  $c$ is the
speed of light, past which its signal is assumed to be so weak that
its signal-to-noise ratio falls below that which is detectable under
reasonable assumptions.

There are two possible scenarios for the broadcast to play out. The
first is that $Lc > D$ whereby the CC's lifetime is so long that its
signals reach their maximum distance even while the CC is still
actively broadcasting. Second, is $Lc < D$ where the maximum
distance of the signal is reached sometime after the CC has already
become extinct and ceased actively broadcasting.

The goal will be to assess the total volume-time filled by the
signal over its broadcasting period and combining this with the
Drake equation to estimate the probability that a signal of
electromagnetic radiation from one CC will be detectable by another
CC. If there is a high probability that a CC should exist within the
volume of space occupied by a broadcasted signal, then contact is
likely. If not, then the signal is considered unheard. It is also
assumed that no other factors such as colonization or interstellar
travel are present. This can be calculated using only the variables
in the Drake equation with the addition of $D$. To calculate the
volume-time swept by the signal we integrate the total volume the
signal saturates at a given time by the time it broadcasts

\begin{equation}
V*t = \int^{t_f}_{t_0}V(t) dt
\label{volumetime}
\end{equation}

\begin{figure}

    \centering
    \includegraphics[height=5in, width=6in]{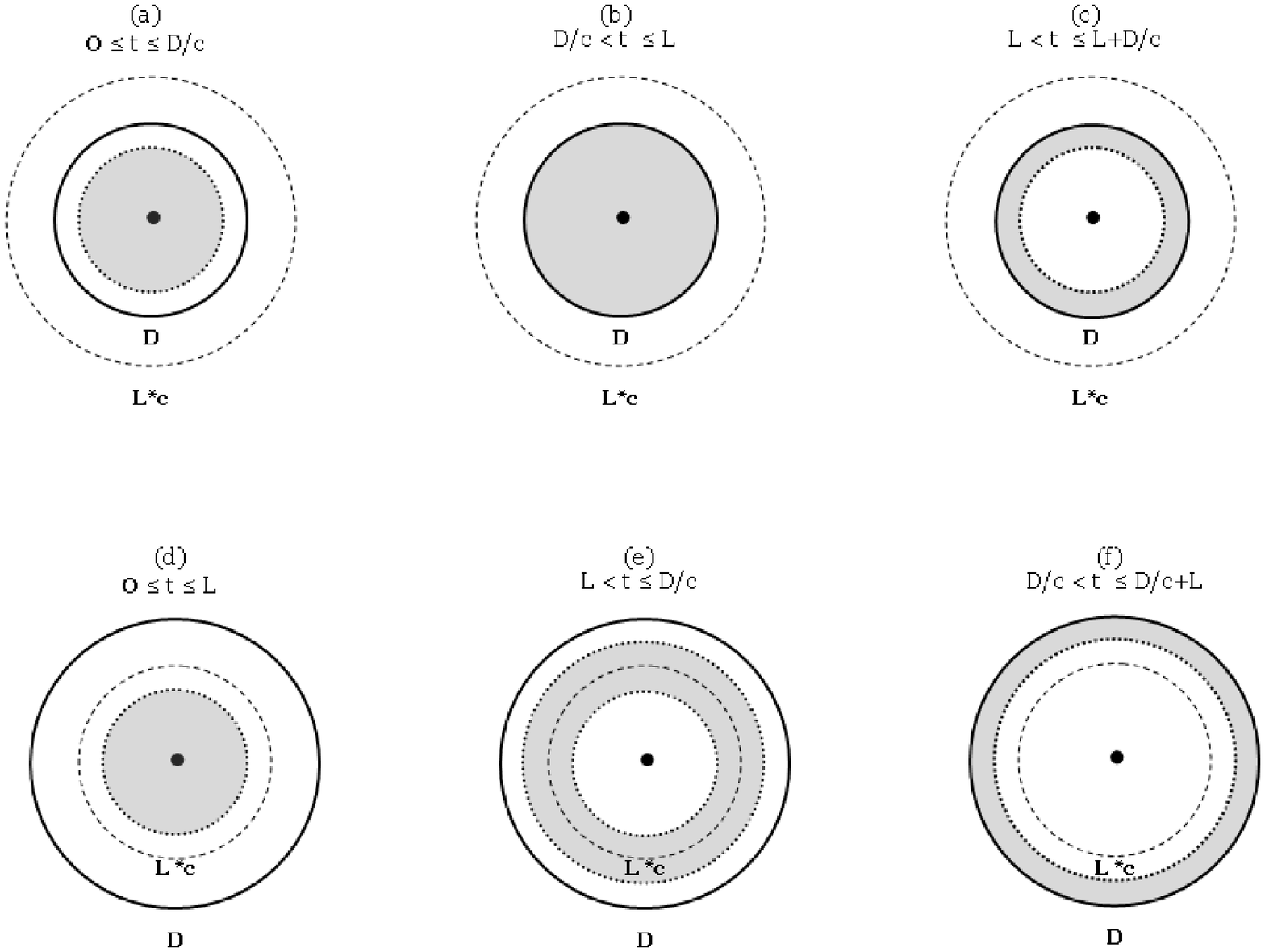}
        \caption{Areas of coverage for the propagation of an expanding signal. (a)-(c) assumes $D<Lc$, (d)-(f) assumes $D>Lc$. The shaded portion is the relative area being swept by the signal.}
    \label{lightcones}
\end{figure}

In figure \ref{lightcones} the volumes filled by the signal are
shown for three points in time under the two possible relationships
between $Lc$ and $D$ explained above. Table \ref{vteqtable} states
$V*t$ under each situation.

\begin{table*}[!t] \vspace{1.5ex}
\begin{tabular}{|c|c|}
\hline \label{vteqtable} (a)&$\frac{1}{3} \pi \frac{D^4}{c}
$\\

(b)&$ \frac{4}{3} \pi D^{3}L - \frac{4}{3}\pi\frac{D^4}{c}
$\\

(c)&$ \pi\frac{D^4}{c}$\\

(d)&$ \frac{1}{3} \pi L^{4}c^3
$\\

(e)&$ \frac{4}{3} \pi c^{3}L
 \left[\frac{D^3}{c^3}-\frac{3}{2}\frac{D^{2}L}{c^2} +
\frac{L^{2}D}{c}-\frac{1}{2}L^3 \right]
$\\

(f)&$ \frac{4}{3} \pi D^{3}L - \frac{1}{3} \pi L^{4}c^3
$\\
\hline
\end{tabular}
\caption{The equations corresponding to the different expanding
coverages of the signals in figure \ref{lightcones}. Note the first
terms in (a) and (d) correspond to the lightcone from the origin to
$D$ and $L$ respectively.}
\label{vteqtable}
\end{table*}

In table \ref{vteqtable} are the equations for $V*t$ for images a-f
in figure \ref{lightcones}. Both sets of equations give the same
value in the case $Lc = D$.

Now in order to integrate these with the Drake equation, we modify
the equation to calculate the average CC density. This is the
expected number of CCs per unit space. We do this by changing the
term, $R^*$ to become $r^*$, the stellar creation rate per unit
volume. Therefore, we can calculate the expected communicative
civilization density as

\begin{equation}
n = r^{*}f_{s}L
\label{drakedensity}
\end{equation}

However, when trying to calculate the number of CCs that can detect
a signal, we replace $L$ with the volume-time to get

\begin{equation}
N = r^{*}f_{s}\sum^{t_{f}}_0 V*t
\label{contactcount}
\end{equation}

where a contact is made with almost certainty if $N\geq 1$.
Volume-time is represented by the following for $Lc > D$ and $Lc <
D$

\begin{equation}
\frac{1}{3} \pi \frac{D^4}{c} + \frac{4}{3} \pi D^{3}L -
\frac{4}{3}\pi\frac{D^4}{c} + \pi \frac{D^4}{c}
\end{equation}

\begin{equation}
\frac{1}{3} \pi L^{4}c^3+\frac{4}{3} \pi c^{3}L
 \left[\frac{D^3}{c^3}-\frac{3}{2}\frac{D^{2}L}{c^2} +
\frac{L^{2}D}{c}-\frac{1}{2}L^3 \right] + \frac{4}{3} \pi D^{3}L -
\frac{1}{3} \pi L^{4}c^3
\end{equation}

The first and final terms in each equation cancel giving

\begin{equation}
\frac{4}{3} \pi D^{3}L
\end{equation}

\begin{equation}
\frac{2}{3} \pi L
 \left[4D^3-3D^2Lc+2L^2Dc^2-c^3L^3\right]
\end{equation}

\section{Communicative civilization density and contact}

Taking equations \ref{drakedensity} and \ref{contactcount} into
account, the question naturally arises of how the density of CC
affects the probability of contact. Solving for $r^{*} f_{s}$ in
\ref{drakedensity}, two variables that are otherwise extremely
difficult to estimate, and changing equation \ref{contactcount} we
can see that

\begin{equation}
\label{cooldrake} N = \frac{n}{L}\sum^{t_{f}}_0 V*t
\end{equation}

We now have a probability of a signal being detected by another CC
only in terms of the average CC density in the region considered,
the average lifetime of a CC, and the volume-time which implicitly
incorporates the maximum distance, $D$. For $Lc > D$ and $Lc < D$,
$N$ follows as
\begin{equation}
N = \frac{4}{3}\pi n D^3
\end{equation}

\begin{equation}
N = \frac{2}{3}\pi n
\left[4D^3-3D^2Lc+2L^2Dc^2-c^3L^3\right]
\end{equation}

In the first case of $Lc > D$ the number of CCs contacted depends only on the signal horizon distance and is independent of the lifetime of the civilization. In short, it is the expected number of CCs in the volume of space specified by the signal horizon distance. Where $Lc < D$, there is a more complex relationship where the broadcasting CC lifetime is an important factor. Given a threshold of $N=1$, we can thus derive
any of the variables if two others are known. In particular, given
assumptions on the average lifetime of CCs and the average
detectable distance of their signals, we can estimate the minimum CC
density and for a given volume, the minimum number of CCs necessary
to make communication probable.

\begin{figure}

    \centering
    \includegraphics[height=3in, width=3in]{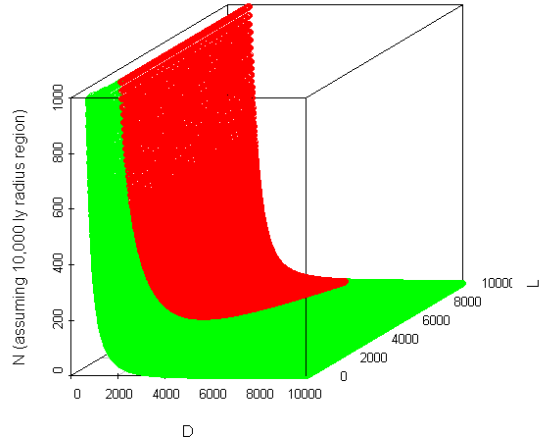}

        \caption{A graph of the minimum number of civilizations needed in a volume of radius 10,000 ly to make both contact (green and red area) and reciprocal contact (red area) likely based on $L$ and $D$.}
    \label{civplot}
\end{figure}

Using a basic measure of our galactic neighborhood as a sphere with
a radius of 10,000 light years, figure \ref{civplot} shows the
number of CCs necessary for the CC density to be high enough to make
contact likely within this volume. This assumes stars, habitable
conditions, and CCs are distributed isotropically which is unlikely
at best but a starting assumption. As expected, for CCs with very
short lifetimes or very short signal horizons, the CC density must
be massive to expect any contact. However, as CCs last longer and
have signals with more durable range, the minimum density decreases
hyperbolically until for high values of lifetime or signal horizons,
only one other CC is necessary for contact to be probable.

What is most interesting about this analysis is that it demonstrates
it can be possible for many CCs in the same galaxy to never contact
one another. For example, even assuming the average CC has a
lifetime of 1,000 years, ten times longer than Earth has been
broadcasting, and has a signal horizon of 1,000 light-years, you
need a minimum of 750 CCs in the galactic neighborhood to reach a
minimum density. For example, if there were only 200 CCs in our
galactic neighborhood roughly meeting these parameters,
probabilistically they will never be aware of each other. The
restriction of $D$ of 1,000 light-years is probably too conservative
since under our current technology, the Arecibo observatory can
detect a signal of its same broadcasting strength at 1,000
light-years. For the same $L$, increasing $D$ to 2,000 reduces the
minimum number to 65 and increasing $D$ to 5,000 reduces it down to
3, though the signals would likely be received after the demise of
the broadcasting CC.

These findings can give pause to both those who predict no other CCs
or those who predict a high number of CCs in our galactic
neighborhood. Arguing that the lack of contact signifies the lack of
CCs may be tempered with the fact that if there is a signal horizon,
even a galaxy replete with life may have relatively isolated CCs in
the absence of interstellar travel or extremely power signals. On
the other hand, high estimates of CCs in our galactic neighborhood
does not guarantee that there will ever be contact between them,
especially reciprocal.

Reciprocal contact can be estimated. On average, the signal needs to
reach a volume of space of $1/n$ to reach another CC. The average
distance between CCs and the average time to reach another CC can be
derived from the two equations

\begin{equation}
\frac{1}{n} = \frac{4}{3}\pi(r)^3
\label{mindist}
\end{equation}

\begin{equation}
\frac{1}{n} = \frac{4}{3}\pi(ct)^3
\end{equation}

In \cite{ETsmart} Duric and Field also calculate the average density
of civilizations based on a time available for intelligible contact
and come up with an average distance between CCs quantitatively
similar to equation \ref{mindist} where $r \approx
\frac{1}{n^{1/3}}$. If $2t \leq L$ then reciprocal contact is likely
on average assuming a CC immediately detects a signal on arrival and
immediately sends a response back which arrives before the first CC
dies out. This predicts a broadly social universe where, much like
the old American public television children's show Mr. Rogers'
Neighborhood, everyone knows their neighbor. However, in time $t$
both CCs could be aware of each other though they have not yet
communicated.

\subsection{Estimating D}
Estimations of $D$ are necessarily bound to considerations
of broadcasting and receiver abilities that would dictate an average
distance for intentionally or unintentionally broadcast signals to
be received. A proper estimate of $D$ would necessarily depend on
the broadcast frequency, bandwidth of the signal, broadcast power,
and both the transmitting and receiving antenna sizes. In
\cite{kuiper}, a range equation for a signal is presented of the
form

\begin{equation}
\frac{\eta_T\eta_Rd_t^2d_r^2P_t}{\lambda^2R^2}=2\times10^3k_Dk_RT\left(\frac{B}{\tau}\right)^{1/2}
\label{range}
\end{equation}
where $\eta_T$ and $\eta_R$ are the aperture efficiencies
of the transmitting and receiving antennas with diameters of $d_t$
and $d_r$ respectively. $P_t$ is the radiated transmitter broadcast
power (megawatts), $\lambda$ is the broadcast wavelength (meters),
$R$ is the maximum range (light-years), $k_D$ and $k_R$ are the
detection efficiency and receiver efficiency factors, $T$ is the
system noise (kelvin), $B$ is the bandwidth (hertz), and $\tau$ is
the integration time needed to detect the signal (seconds). By
estimating some of the constants and and realizing the maximum
$\tau$ is $L$, you can derive a minimum necessary broadcast power
for a given $D$ or estimate $D$ for an assumed broadcast power. You
can also estimate upper limits for bandwidth and minimum antenna
sizes. For example, using the estimates for the variables in
\cite{kuiper} of $\eta_T = \eta_R$=0.6, $k_D$=5,
$k_R$=2,$T=20^\circ$K, $d_r=d_t$=26 m and assuming a bandwidth of 1
kHz at a $\lambda$ of 12 cm for $D$ of 1,000 ly and $L$ of 1,000
years we would need a minimum broadcast power of 6.2 MW, assuming
integration time is the full life of the CC, which is probably far
too large of an estimate.

\section{Conclusion}

The Drake Equation, almost five decades after its debut, remains as
controversial and inconclusive as it was at its inception. A good
viewpoint is raised in \cite{drakeoutdate} that discusses the
problem that the Drake equation should not be viewed as an exact
mathematical equation, especially since it includes both
astronomical variables such as star formation rate and planet
abundance, biological variables such as the probability for the
emergence of life, as well as more socioeconomic variables such as
the probability for the development of civilization and advanced
communication. There is no way to find a `right' value for these
variables.

However, the search for extraterrestrial life need not suffer
because of no immediate obvious contact despite high estimates of
life in our galaxy and universe. Under appropriate considerations,
if the density of life is lower than a certain threshold and
assuming colonization driven contact is unlikely, communicating
civilizations could remain completely ignorant of each other. Of
course these strict constraints can be circumvented by interstellar
travel or permanent automatic beacons. This paper has attempted to
add some additional considerations to the Drake equation which will
allow us to more feasibly estimate the conditions under which we can
hope to eventually discover we are not alone.


\begin{thebibliography}{1}

\bibitem{fermi}The Fermi Paradox is based on a possibly apocryphal
question Enrico Fermi supposedly asked at Los Alamos in the 1940s:
given that it seems there is a high probability for the evolution of
intelligent life, why haven't we found it?
\bibitem{drake}Drake, FD, Discussion at Space Science Board, National Academy of Sciences Conference on Extraterrestrial Intelligent Life, (1961)
\bibitem{sagan}Shklovskii, IS \& Sagan, C, \emph{Intelligent life in the
universe}, New York: Dell, (1966)
\bibitem{drakeadvanced}Kreifeldt, JG, ``A formulation for the number of communicative
civilizations in the galaxy'', Icarus, \textbf{14}, pp.419-430,
(1971)
\bibitem{drakeadvancedb}Wallenhorst, SG, ``The Drake Equation
reexamined'', Quar. J. of the Royal Astron. Soc. , \textbf{22},
pp.380-387, (1981)
\bibitem{saganrebut}Tipler, FJ, ``Extraterrestrial intelligent beings do not
exist'', Quar. J. of the Royal Astron. Soc., \textbf{21}, pp.267-281
(1980)
\bibitem{saganrebutb}Cirkovic, MM, ``The Temporal Aspect of the Drake Equation and
SETI'', Astrobiology, \textbf{4}, pp.225-231 (2004)
\bibitem{ETsmart}Duric, N \& Field, L, ``On the detectability of intelligent civilizations in the galaxy'',
Serb. Astron. J., \textbf{167}, pp.1-10 (2003)
\bibitem{kuiper}Kuiper, TBH \& Morris, M, ``Searching for extraterrestrial
civilizations'', Science, \textbf{196}, pp.616-621 (1977)
\bibitem{drakeoutdate}Burchell, MJ, ``W(h)ither the Drake
Equation?'', Intl. J. of Astrobiology, \textbf{5}, pp.243-250,
(2006)


\end{thebibliography}
\end{document}